\documentclass{article}
\begin{document}
\title{On the Paradoxical Book of Bell}
\author{{Marek \.Zukowski}\\
{\small \protect{Institut Fizyki Teoretycznej i Astrofizyki, Uniwesytet Gda\'nski, 80-952 Gda\'nsk, Poland}}}
\maketitle
%\begin{document}
%\maketitle
%\doublespacing
%\begin{abstract}
%{This is an essay-review on a recently  re-issued book of John Bell {\em Speakable and Unspeakable in Quantum Mechanics}. The discussion concentrates %around the Bell Theorem, its assumptions, consequences and frequent overinterpretations. }
%\end{abstract}

%Keywords: {\em Bell, Bell Theorem, Quantum Theory}

%\pacs{03.65.Ud}

%\maketitle

\section{Introduction}
The famous book of Bell, {\em Speakable and Unspeakable in Quantum Mechanics}, has been re-released last year (Bell, 2004). I feel extremely honored to be asked by {\em Studies} to write an essay-review on it. The works of Bell had an enormous influence on my scientific life. The books covers quite a range of topics, but undoubtedly everyone would agree that it centers around the most outstanding input of Bell to science, that is His Theorem. Since my intuition tells me that the Editors chose my humble person to write an essay-review on the book because of my continuing interest in the Theorem, the review would also concentrate on the chapters in the book devoted to this topic. I shall try to show the paradoxical aspects of Bell's research on the Einstein-Podolsky-Rosen paradox, as well as the paradoxical history of the paradox which emerges from the book.

\section{Einstein-Podolsky-Rosen paradox  before Bell}

The EPR paper, Einstein, Podolsky and Rosen (1935), was a shock to Bohr and his Copenhagen Interpretation followers. In modern terminology the paper suggests that 
perfect quantum correlations, which are possible for two sub-systems in an  entangled state can be used to define a missing element in quantum theory, namely "elements of reality". Such elements of reality would, according to EPR, exist because measurements on one sub-system, $S_1$, when the two subsystems are far far away from each other, cannot directly influence the other system, $S_2$, while they can be used to determine (at-a-distance) the results of possible measurements of those variables of $S_2$, which are perfectly correlated with the measured variables of $S_1$. Therefore a certain property of system $S_2$ seems to be (perfectly) predictable without actually measuring it, thus it seems to be possessed by the sub-system. Moreover, using the fact that the observer (since 1990's, Alice), who makes measurements on system $S_1$ is {\em free} to choose the variable  she wishes to measure,
EPR argue that in principle she could choose between complementary observables. Since the entangled state studied by EPR (as well as the modern
maximally entangled pure states) reveals correlations in complementary observables of the two subsystems, values of two complementary variables of system $S_2$ are in principle determinable-at-a-distance, without in any way disturbing the system $S_2$. Thus they must be real properties possessed by the subsystem $S_2$. This is of course in a direct contradiction with Heisenberg's uncertainty (which applies to pairs of complementary variables),   and suggests that quantum description is incomplete (note that what was challenged was not validity of the quantum formalism, but its completeness).

The response of Bohr (1935) did not try to find a direct inconsistency of EPR's ideas with quantum predictions. Instead, he tried to expose the counterfactual elements in EPR reasoning. Alice's choice of the variable to be measured affects the very conditions on which her deterministic prediction-at-a-distance of the property of the other system is based. In modern orthodox wording this can be, I hope, put as follows. Her choice of the measured variable, and the outcome of her measurement, naturally, changes her state of knowledge on both correllated systems. The changed state of knowledge leads to new predictions, which can be deterministic, due to the perfect correlations \footnote{Note that Bob, her partner, who can make the measurements on $S_2$, knows nothing about her choice of the measurement, and her result; for him, without getting a ``phone call'' from Alice, the state of his knowledge does {\em \bf not} change with Alice's measurement.}. Since she cannot determine the other complementary variable of her sub-system, the complementary variable of Bob's system remains to her totally undetermined, because the conditions for her predictions changed in such a way that the information on complementary variables has been irreversibly erased.

Paradoxically at that time existed a, stupid according to Bell, see Mermin (1993), proof of the impossibility of a hidden variable extension of quantum mechanics by von Neumann  (1932). A counterexample to this proof was found by Bohm (1952), who  also put the EPR paradox in a form involving two maximally entangled spins $1/2$ (Bohm, 1951), which was used by Bell (1964) in his trail-blazing paper. Surprisingly, neither EPR nor Bohm analyzed carefully the proof of von Neumann. The existing confusion: the EPR suggestion of the missing elements of reality (hidden variables?), Bohm's model giving a causal-like hidden variable completion of quantum mechanics, versus Bohr's verbal refutation of the EPR argument and the no-go for hidden variables theorem of von Neumann, waited for Bell to be clarified in his two brilliant papers.

\section{Enters Bell}

Well, when I got first involved in topics related to ``foundations of quantum mechanics", late 1980s, it was indeed very difficult to publish papers on such subjects in non-fringe journals. All professors shouted: ``shut up and calculate". In the earlier times, it must have been even worse.
Definitely not too many people around to discuss such topics. Not surprisingly, form a sociological point of view, the discussion about the foundations of quantum mechanics was continued almost only by the dissidents of the ruling orthodoxy. Bell was one of them. To him quantum mechanics was ``rotten".
But fortunately the beautiful aspect of the real science is the scientific method. As long as it is not abused, progress can be made, no matter what are the beliefs, hopes or conjectures of those involved (the opening paragraphs of chapter 18 of Bell's book offer his  reflections on this theme). Judging from the chapter ``On the impossible pilot wave" Bell started to hope for a consistent removal of such notions as observer, measurement, wave packet collapse, from the quantum language after reading the paper of Bohm (1952). The fact that he formulated his theorem as late as in early 1960's shows to what extent the environment was discouraging this type of research. However, he kept on thinking...

In two papers, as it should be, paradoxically published in a `time reversed' sequence, Bell clarified all that confusion. The earlier paper, Bell (1966), published after the later one, Bell (1964), demolished von Neumann's assumptions on the structure of hidden variable theories. These, under the rigorous scrutiny of Bell, turned out to be out of touch with anything that could be reasonably requested form an underlying classical statistical theory of quantum phenomena. Von Neumann's theorem turned out to be totally irrelevant.

The next paradox about the ``earlier" paper is that, right after the refutation of von Neumann's theorem Bell proves... something that is, or was, commonly known as the Kochen-Specker theorem. The paradox is in the fact that he publishes his proof one year before Kochen and Specker (1967), and since this is his ``earlier" paper, his proof must have been ready at least two more years earlier. The Kochen-Specker theorem of Bell refutes, basing on a lemma by Gleason (1957), the so-called non-contextual hidden variable theories, for three or more dimensional quantum systems.
Such theories assume that for an individual system, belonging to a statistical ensemble of equivalently prepared quantum systems (i.e. described by a certain state), one can introduce predetermined values of all possible observables. For a specific observable its predetermined value is context-independent, in other words, does not depend on the possible other observables, which are, or can be, co-measured with it. Since only observables with a degenerate spectrum can belong to two different complete sets of commuting observables, Kochen-Specker theorem holds for systems which are at least three dimensional (for a two dimensional system an observable with a degenerate spectrum is simply a constant).

In the last chapter of the ``earlier" paper one finds a discussion of Bohm's hidden variable model for two spin $1/2$ particles. Bell shows elegantly that the model is non-local: ``disposition of one piece of [measuring] apparatus affects the results obtained with a distant piece. In fact the Einstein-Podolsky-Rosen paradox is resolved in the way which Einstein would have liked least" \footnote{The non-locality of the model was known to Bohm.}. In the final paragraph Bell stresses that ``there is no {\em proof} that any hidden variable account of quantum mechanics {\em must} have this extraordinary character". In footnote 19 he adds: ``since the completion of this paper such a proof has been found", and he cites his 1964, `` later" paper, erroneously as J.S. Bell, {\em Physics} {\bf{1}}, 195 (1965), a mistake to be followed by many \footnote{Among those citing the Bell's paper the majority of citations of over 1900 prefers 1964 as the official publication year, over 90 votes for 1965, and in one paper the choice is 1963...}.   
  
Note that the ``earlier" paper contained all ingredients to formulate Bell's Theorem: discussion of non-contextual hidden variable theorems, and a discussion of non-locality of an existing hidden variable theory that perfectly reproduced quantum predictions. In the ``later" paper Bell analyzes the EPR problem in Bohm's version: the context here for Bob is the experiment performed by Alice, very far away and at the same moment (for a certain relativistic observer), thus Einstein's causality prohibits the context to influence the actions of Bob, and the local results of his simultaneous measurements.

\subsection{The Theorem of Bell}

The surprising fact about Bell's book is that somebody who wants to {\it learn} the essence of Bell's theorem should not read the chapter two, which contains the aforementioned ``later", most famous, paper. The paper is indeed a trail-blazing one, but as a matter of fact can be treated as comment on the paper of EPR. It addresses directly the problem of elements of reality, called by Bell ``predetermined measurements", via the perfect correlations, which are the property of a singlet state of two spins $1/2$. The local hidden variable construction behind the {\em Bell-type} inequalities is introduced, without any need for these perfect correlations, in chapter 4, in which the Clauser-Horne-Shimony-Holt (1969) (CHSH) Bell-type inequality is re-derived. However, still not all assumptions are spelled out. All that shows, that our, and Bell's understanding of his inequalities was evolving. The full exposition of the assumptions begins on p. 150, in the classic chapter/article on ``Bertlmann's socks". With today's wording these assumptions boil down to
\begin{itemize}
\item
{\em Realism}: to put it short: results of unperformed measurements have certain, unknown, but fixed, values. In Bell wording this is equivalent to the hypothesis of the existence of hidden variables.
\item
{\em Locality}: ``the direct causes (and effects) of events are near by, and even indirect causes (and effects) are no further away then permitted by the velocity of light" (p. 239), in short, events and actions in  Alice's lab cannot influence directly simultaneous events in Bob's lab and his acts, etc. 
\item
{\em ``Free will"}: the settings of local apparata are independent of the hidden variables (which determine the local results), and can be changed without changing the distribution of local hidden variables (p. 154), in short, Alice and Bob have a free will to fix the local apparatus settings, or more mildly, one can always have a stochastic process that governs the local choices of the settings, which is statistically independent from other processes in the experiment (especially those fixing the hidden variables).
\end{itemize}

It must be stressed that with the use of the above assumptions {\em one cannot derive the original Bell inequality}. It was derived using additionally the specific property of the singlet state, namely perfect anti-correlations of measurement results for two distant observers using identical settings of their apparata. Authors who are not aware of this fact, and its consequences, are often lead to exciting, but meaningless results, like that there are separable quantum states that violate the original inequalities, or that the original Bell inequalities can be violated more strongly than the CHSH ones \footnote{The critical (data-flip) threshold maximal error rate for a violation of the CHSH inequalities, when the singlet correlations are observed, is $14.6\%$, whereas in the case of the original ineqality it seems, but only seems to be $16.6\%$}, etc. In the first case the lack of perfect EPR correlations leads to the violation, whereas in the second case the very existence of errors destroys perfect correlations, and therefore one cannot use the original inequality to analyze the data. A newcomer to the field should be strongly discouraged to read (only) the original paper of Bell as an introduction to the subject of Bell inequalities. He/she has the read the whole book.

With the assumptions listed above, one can derive the CHSH inequalities and all other ones, including those which are generalizations to the problems 
involving more than two parties. In order to avoid confusion, only  inequalities based on these three assumptions, should be called  Bell-type inequalities. Only then they are general necessary conditions for local realism\footnote{The original ones are necessary conditions for local realism {\em and} specific perfect correlations.}. We have yet another paradox: the original Bell inequality, using this convention, is not an inequality of the Bell-type. With the modern terminology Bell-type inequalities define the faces of the polytope of all possible local realistic correlations. The original Bell inequalities cut the polytope into two pieces. Further, because they rest upon the perfect correlations, they cannot be used to analyze any experimental data, because of the inevitable imperfections of these.

One more paradox associated with the Bell Theorem: it took 25 years before  systematic investigations of three or more particle experiments began! Most probably the (wrong) rule of thumb worked: the more particles are involved, the more classical is the system. So nobody expected anything exciting to happen for three or more particles. It looked like that everything what was to be said about Bell's theorem was already said. The 1989 paper of Greenberger, Horne and Zeilinger (1989) (GHZ)  changed all this. The superpositions due to entanglement of many particles lead to even more drastic versions of Bell's theorem than the two particle superpositions. The GHZ paper started the second phase of Bell's revolution: an intensive search for new phenomena a involving quantum entanglement, for many particle systems or for non-spin observables. This phase continues and involves physicists preaching all kinds of interpretations of quantum mechanics, including the orthodoxy. 

The third phase was just around the corner. The paper of Ekert (1991) showed that Bell-EPR correlations can be used for distribution of a cryptographic key, and Bell's inequalities could be used to estimate the level of security the quantum cryptographic protocol. Then came the teleportation
 paper, Bennett et al (1993),  which involved Bell states (that is, maximally entangled two qubit systems) as the quantum channel of the process, and Bell-state measurements, as source of the classical information to be transfered, and at the same time erasers of the state to be teleported. A new branch of physics started to emerge: quantum information science. Basically all books on quantum information start with Bell's theorem, thus the 1964 article of Bell can be now viewed as the first paper on quantum information. The debate, which was initially of almost a philosophical nature, to the surprise of the physics community, turned out to be of key (not only cryptographic) importance in the development of the {\em direct} applications of quantum phenomena in quantum communication and quantum computation. The word {\em direct} is very important here: quantum phenomena are not used to make possible classical communication and computation schemes, but rather to construct new schemes of an essentially quantum nature, which  employ the paradoxical nature of the quantum realm, to achieve tasks unachievable by classical means.

   \section{Overinterpretations of Bell's Theorem}
   
The consequence of violation of Bell's inequalities is that either one of the listed assumptions is wrong, or two of them or all. The free will assumption, especially in its soft form of statistical independence, seems highly unlikely to be wrong. It is very difficult to imagine a clockwork world which is tuned exactly in such a way, that whenever a Bell experiment is made, the intercorrelations of the entire system of the source, the two measuring apparata, and the pseudo (inevitably, in this case) random number generators which decide the local settings, are invariantly  such that the required statistical independence is never ever achievable. Note, that in classical probability one can always assume underlying determinism, but nevertheless the formal definition of independent events is not void. Therefore the trouble seems to be with the other two assumptions. And here comes yet another paradox: the consequences of Bell's theorem as they are now most frequently presented to the entire physics community.

Let us look at the ruling classification ruler: Physics and Astronomy Classification Scheme (PACS). Go to the web page, type  into {\em find a string}
the word {\em nonlocality}, and what pops up? Here it is: {\em PACS 03.65.Ud Entanglement and quantum nonlocality (e.g. EPR paradox, Bell's inequalities, GHZ states, etc.)}. That is, according to the authoritative source, Bell's theorem in its various versions is linked with ``quantum nonlocality". As quantum nonlocality is not an assumption of Bell Theorem\footnote{It seems to be useful to use the following distinction: Bell's inequalities hold for local realistic systems, and Bell's Theorem states that they are violated by quantum systems. That is Bell's inequalities are not synonymous with  Bell's Theorem}, therefore we are led to the conclusion that it is its consequence. As the source is very authoritative, then we are lead to yet another conclusion: it is its the inevitable consequence! Of course we started our search with trivial nonlocality, and now we have `quantum nonlocality' which may be a different beast (note that there is a significant difference between a chair and an 
{\em electric} chair).

There is a plethora of papers that use phrases like ``the proof of nonlocality", ``violations of Bell inequalities, that is quantum nonlocality", 
or ``quantum nonlocality, that is violations of Bell's inequalities", etc. It is very difficult (for me) to understand why some people jump to this conclusion. There is nothing in the quantum formalism that would necessarily imply non-locality. One may think that the state collapse postulate suggests it. But states are predictive tools for ensembles of equivalently prepared systems\footnote{With such a statement, I hope, do agree all those who use, or think about, quantum mechanics. One may demand from the state to be more than this predictive tool, but this does not harm its predictive powers. To put it short, Born's rule is applied by everyone.}. Once an additional  measurement is made, which splits the ensemble into physically distinct subensembles, the predictions must change when one moves form one subensemble to another. Additionally the interaction with the measuring device must have its effects: quantum mechanics tells us that we get new information only via interaction. Closed systems follow a deterministic (unitary) evolution - collapse never occurs. The states are themselves unmeasurable for single systems, i.e. they cannot be not their property\footnote{For a single system there is no measurement procedure that determines its state, prior to this measurement. This is equivalent to the no-cloning principle.}. Therefore their instanteneous collapse, upon a measurement of some observable, cannot lead to any observable non-local consequences. And indeed we have the non-signaling principle. Also in electrodynamics one can use as auxiliary fields the potentials. They can be put in the Coulomb gauge, which is non-relativistic, but still the observable phenomena follow Einstein's causality\footnote{The Aharonov-Bohm effect is essentially of a (magneto) static nature. One cannot use it for faster than light signaling.}.

Realism is neither an assumption of quantum mechanics, nor can it be derived from it (despite the hopes of EPR), without violating Einstein's causality (Bell's Theorem!).
It is in a clear cut contradiction with the principle of complementarity. In quantum theory the values of two noncommeasurable observables cannot be simultaneously defined, even only in theory. Generally one cannot construct, within the quantum theory, a joint (nonnegative) probability distribution for two complementary variables.  
In an experiment we can define whether the given photon is horizontally or vertically polarized, but its circular polarization is neither a stochastic mixture of right and left hand ones, nor does it have a hidden value. It is simply undefined, i.e. for quantum formalism it does not exist.  
However, from a realistic point of view it is merely unknown, but does exist, or, equivalently, is predetermined by some fixed hidden variables. This in turn implies the existence of a joint probability distribution of horizontal/vertical and left/right circular polarizations.\footnote{Paradoxically such a distribution in not possible for (entangled) pairs, or more photons, but can be shown to exits for a single photon (see chapter 2).}

It is an interesting fact, that the shrinking majority, or hopefully already a shrinking minority, of physicist who think the Bell's theorem is irrelevant for physics, usually react very aggressively when somebody, trying to convince them that Bell's theorem is interesting, tells them that they should manipulate with results of two measurements of noncommeasurable observables, and treat them on equal footing, as two unknown but fixed values. That is, they oppose very strongly the assumption of realism, to such an extent, that they do not want to listen to any further element of the derivation of Bell's inequalities. Surprisingly the, hopefully one day shrinking, majority out of those who accept Bell's theorem as a great achievement of science, find violations of locality as as its consequence. Or, is it just marketing? 

It may be the case that we can indeed talk about quantum nonlocality, but definitely it is not a straightforward consequence of Bell's theorem.
Bell himself uses the term quantum mechanical nonlocality (the title of chapter, of course, 13). However, as he was a realist, for him such was the consequence of his Theorem. He was very interested in Bohm's formulation of quantum mechanics. Later he flirted with the Ghirardi, Rimini and Weber (1986) spontaneous localization theory (which is a modification of quantum mechanics, not just its interpretation).

He uses the term quantum mechanical nonlocality, but in many of the chapters of the book one can find the full list of possible consequences of His Theorem, including that `` it may be that Bohr's intuition was right - in that there is no reality below some `classical' macroscopic level". But he, the skeptic, immediately adds ``then the fundamental physical theory would remain fundamentally vague... ".

The new edition of the book contains two `bonus' chapters, which were written after the first edition appeared. Unfortunately in the last chapter Bell suddenly calls the CHSH inequality `locality inequality' (p. 245). Well, perhaps here begins the history of PACS 03.65ud (ud = ugly deceptive?). Or is it just a vague statement overlooked in proofreading, or finally a Freudian slip? 

\section{Final remarks}
Bell's Theorem is one of the greatest discoveries of modern science, of direct and far reaching philosophical consequences. One cannot imagine any comprehensive discussion about the interpretation or foundations of quantum mechanics that ignores this input of Bell. All quantum information textbooks begin with it, and there is even a conjecture that only quantum processes that violate some (generalized) Bell inequalities can find quantum informational applications, see Scarani and Gisin (2001).
It influences also the voodoo science: attempts to prove Bell's Theorem wrong appear with surprising regularity, and put the old tasks of building the perpetuum mobile, or demolishing of special relativity, down in the list of priorities.
Thus an extreme care must be taken when the theorem is presented. On the one hand, one cannot say too little about the assumptions, leave some of them tacit. On the other hand, one must avoid overinterpretations of its implications.

Will the re-issue of the book of Bell be helpful for us in this respect? Yes, indeed! But only provided it is read back-to-back, and carefully. Paradoxically the greatest blow to the realistic interpretation of quantum mechanics was delivered by one of its advocates. He wiped out an enormously large class of them: the most intuitively acceptable class - the local realistic interpretations. This is a triumph of the scientific method, over the individual preferences of the scientist. Nevertheless, the book must be read, having in mind that the whole story is presented by somebody holding strongly nonorthodox views, with the obvious bias of the Freudian slips.  This is why this essay tries to stick to the orthodox point of view (which is, I must confess, a strong prejudice of the author).

Finally, I must stress that the value of the book is enhanced by the fact that now it starts with the introduction of Alain Aspect, one of the experimentalist who contributed most to the public awareness of the importance of the Theorem. It is in itself a very interesting reading. One must remember, that in the hands of experimenters is the final fate of Bell's theorem. We still wait for a loophole-free Bell experiment. Bell himself was one of those realists who accepted the verdict of the nonprefect experiments, as a falsification of local realism. He kept repeating that if imperfect experiments follow so well the quantum predictions, there is very little room for hoping that perfect experiments would show very strong deviations from quantum theory. Something very orthodox-like indeed!

\section*{Acknowledgements}

Author is supported by an FNP Professorial Subsidy, and MNiI Grant
1 P03B 04927. The work is dedicated to AZ, one of the first orthodox people who noticed the richness of the implications of Bell's Theorem, on the occasion... he knows which.

  \section*{References}
  
\noindent
Bell, J.S. (1964). On the Einstein-Podolsky-Rosen paradox. {\em Physics} {\bf 1}, 195-200.\\

\noindent
Bell, J. S. (1966). On the problem of hidden variables in quantum mechanics. {\em Rev. Mod. Phys.} {\bf 38}, 447-452.\\

\noindent
Bell, J.S. (2004), {\em Speakable and Unspeakable in Quantum Mechanics} (University Press, Cambridge), first published 1987.\\

\noindent
Bennett, C. H., G. Brassard, C. Creapeau, D. Jozsa, A. Peres and W.K. Wootters (1993). Quantum Teleportation. {\em Phys. Rev. Lett.} {\bf 70}, 1895-1898.\\

\noindent
Bohm, D., (1952).A suggested interpretation of the quantum theory in terms of `hidden variables'. {\em Phys. Rev.} {\bf 85}, 166-179; {\em ibidem} {\bf 85} 180-193.\\

\noindent
Bohm, D. (1951). {\em Quantum Theory} (Englewood Cliffs, New Jersey).\\

\noindent
Bohr, N. (1935). Can quantum-mechanical description of physical reality be considered complete? {\em Phys. Rev.} {\bf 48}, 696-702.\\

\noindent
Clauser, J. , M. Horne, A. Shimony, and R. A. Holt (1969). Proposed experiments to test local hidden variable theories. {\em Phys. Rev. Lett.} {\bf 23}, 880-883.\\

\noindent
Einstein, A., B Podolsky and N. Rosen (1935). Can quantum-mechanical description of physical reality be considered complete? {\em Phys. Rev.} {\bf 47} 777-780. \\

\noindent
Ekert, A. K. (1991). Quantum Cryptography based of Bell's theorem. {\em Phys. Rev. Lett.} {\bf 67}, 661-664.\\

\noindent
Ghirardi, G.C., A. Rimini and T. Weber (1986). Unified dynamics for microscopic and macroscopic systems. {\em Phys. Rev.} {\bf D 34}, 470-491.\\

\noindent
Greenberger, D. M., M. Horne and A. Zeilinger (1989).Going beyond Bell's Theorem, in {\em Bell's Theorem, Quantum Theory and Conceptions of the Universe}, edited by M. Kafatos (Kluwer, Dordrecht), p.73.\\

\noindent
Kochen, E. and E. P. Specker (1967). The problem of hidden variables in quantum mechanics. {\em J. Math. Mech.} {\bf 17} 59-87.\\

\noindent
Mermin, N.D. (1993). Hidden variables and the two theorems of John Bell. {\em Rev. Mod. Phys.} {\bf 65}, 803-815.\\

\noindent
Scarani,  V. and N. Gisin (2001). Quantum communication between N partners and Bell's inequalities. {\em Phys. Rev. Lett.} {\bf 87},
 117901.\\

\noindent
von Neumann, J. (1932), {\em Mathematische Grundlagen der Quanten-Mechanik} (Springer, Berlin, 1932), see also {\em Mathematical Foundations of Quantum Mechanics} (Princeton University, Princeton, New Jersey, 1955).\\

\end{document}